# An Evaluation of Visualization Methods for Population Statistics Based on Choropleth Maps

Lonni Besançon, Matthew Cooper, Anders Ynnerman, and Frédéric Vernier

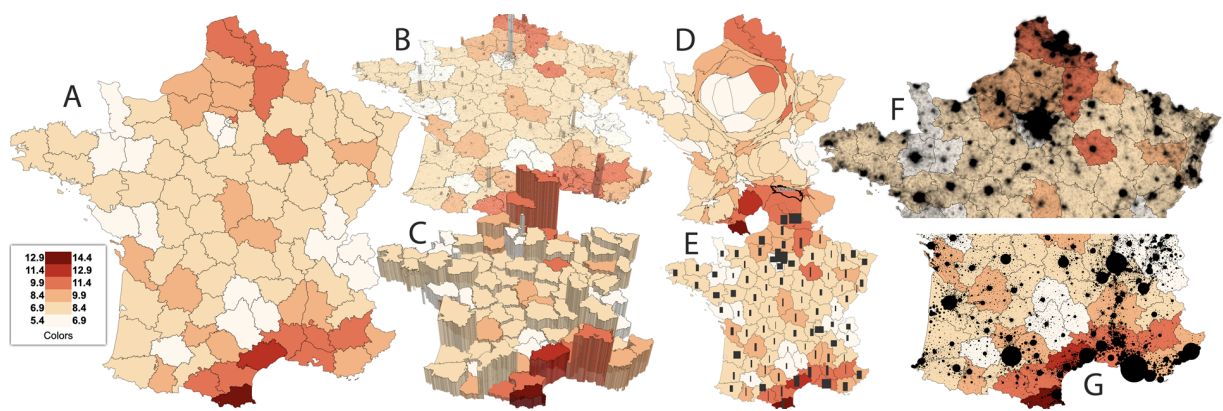

Fig. 1. Choropleth map (A) augmented with 3D extrusion (C), contiguous cartogram (D), and rectangular glyphs (E) at the same level of granularity and with 3D extrusion (B), Heatmap (F) and dot map (G) at a finer level of granularity.

**Abstract**— We evaluate several augmentations to the choropleth map to convey additional information, including glyphs, 3D, cartograms, juxtaposed maps, and shading methods. While choropleth maps are a common method used to represent societal data, with multivariate data they can impede as much as improve understanding. In particular large, low population density regions often dominate the map and can mislead the viewer as to the message conveyed. Our results highlight the potential of 3D choropleth maps as well as the low accuracy of choropleth map tasks with multivariate data. We also introduce and evaluate *popcharts*, four techniques designed to show the density of population at a very fine scale on top of choropleth maps. All the data, results, and scripts are available from osf.io/8rxwg/

✦

## 1 INTRODUCTION

A perennial issue with mapping statistical information onto choropleth maps is that they tend to overemphasize large, yet often sparsely populated, administrative areas because of their strong visual weight [1, 44, 64, 88, 94]. Since choropleth maps usually convey only information about the statistic of interest and no information on the spatial distribution of population, underpopulated areas tend to dominate the map space while the more densely populated are small and hard to distinguish. We hypothesize that large and sparsely populated areas could bias the average information that should be taken from a map visualization and maps could thus be inaccurately interpreted.

There are several solutions to the issue of relative areas [41]. One is juxtaposition: the two variables displayed in adjacent charts. The second is superposition: the two variables encoded simultaneously in a single chart. The last is explicit encoding: directly displaying a single variable computed from the two component variables. Juxtaposition is a viable solution: it clearly allows the user to perceive patterns in each of the two represented variables. However, recent work [22] suggests that integrating the information from two charts is more error prone. Superposition allows the user to understand the influence of both variables. However, bivariate visual representations have been said to be harder to understand [102]. It is thus still unclear which strategies should be adopted to create bivariate maps to properly convey population information in addition to the measured variable usually shown on univariate maps. We aim in this work to address this issue.

The contributions of this work are threefold. We first describe and implement several alternative bivariate map designs displaying a statistic and population distribution. We then report the results of a qualitative evaluation of these designs with 10 visualization researchers and the results of a controlled experiment with 60 participants. Finally, we introduce *popcharts*, designed to display statistical data over a fine-grained population distribution, and report a qualitative evaluation of this approach by the same 10 visualization researchers together with a controlled experiment with 47 participants.

## 2 RELATED WORK

This review considers the design and use of choropleth maps as well as previously conducted studies.

### 2.1 To quantify or not to quantify

The literature has primarily focused on how to design choropleth maps and the problems that each design might exhibit. We briefly sum up past work on discrete and continuous colour coding on maps.

Choropleth maps can be classified into unquantized (i.e., continuous or unclassed, e.g., [107]) and discrete (e.g., [34, 53, 71]) for colour coding. Unclassed maps, initially proposed by Tobler, create unique colours for each values and offer a greater fidelity to the data [17,50,99]. However, they do not account for perceptual limitations of colours [15, 80, 84, 85, 104] and recent work has shown that discrete map could

- L. Besançon is with Linköping University, lonni.besancon@gmail.com.
- M. Cooper and A. Ynnerman are with Linköping University.
- F. Vernier is with Université Paris Saclay and Aviz, Inria.

lead to better performance [75]. We thus decided in this work to discretize the data of the statistic that we wish to show.

## 2.2 Bivariate maps

A common means to display two variables on a map is to use a bivariate colour coding scheme [30, 51, 82, 83, 100], but these have been shown to be hard to interpret [37, 102] and their legend hard to memorize [22]. Consequently, Bivariate colour maps are often limited to a small set of colours [22, 53, 60, 71, 83, 100]. Bivariate colour coding is an integral conjunction: the selective separataton of attention between both attribute encoding is not straightforward. It is therefore not recommended for variables with different units like population on the one hand and a statistic on the other hand to both use colour [87].

Other approaches have focused on augmenting choropleth maps with additional visual elements without relying on an extra colour, starting with Charles Joseph Minard [35] and his attempt to display quantities of meats supplied to Paris by each French department in 1858 [67]. Other maps followed in the 19th century [7, 8]. More recent work has also used this approach with data-dependent glyph augmentations [105] or pixelization/glyph rotations to show uncertainty on a map [60]. Height on a 3D map has also been suggested [49] and tested to observe the evolution of populations [48]. Other approaches rely on a visual property that is closely related to colour, such as opacity [22, 85, 88]. A full taxonomy of possible approaches is described by Elmer [32] who showed that, despite common reluctance to use them, bivariate maps can be successfully interpreted. Elmer tested eight bivariate maps but did not test all possible variants in their taxonomy. We take inspiration from these approaches and incorporate 3D, opacity and glyphs in the set of techniques to display bivariate data. Other studies comparing the representation of geographical data are detailed in Sect. 3.

The study from Correll et al. [22] is particularly interesting for our work. They focused on uncertainty visualization with bivariate maps and compared juxtaposed univariate maps, continuous bivariate maps, discrete bivariate maps and a new version called Value-Suppressing-Uncertainty which aims to reduce the number of categories in bivariate maps. Their studies indicate that continuous colour-based bivariate maps and juxtaposed univariate maps are outperformed by discrete bivariate maps relying on a colour and transparency scheme. We take inspiration from this study but, instead of uncertainty visualization, we focus on population density and merge their results with other work [37, 102] and recommendations [32, 86] to eliminate bivariate colour-coding schemes, preferring colour and transparency [22, 85, 88].

## 2.3 The Benefits of Dasymetric Maps

To avoid the common pitfall of choropleth maps, cartographers have focused on dasymetric maps in which the boundaries of cartographic representation are not arbitrary/administrative but, instead, reflect the spatial distribution of the variable being mapped which may be presented at a fine scale [31]. Many projects have focused on how to derive a fine-scale distribution of the population that is more realistic than in a choropleth map [27, 55, 59, 70, 108, 113]. Researchers have sought out to confirm the benefits of such precisely mapped population densities in different domains [5, 19, 79, 91]. In the case of health outcomes, recent work has shown that choropleth maps tend to lead to either overestimation or underestimation of risk exposure and that finer-population distributions provided by dasymetric maps help [81]. In all mentioned previous approaches, however, fine-scale population distributions are often shown on classically-delimited maps, probably for two reasons. First, the classic divide into counties or other regions is usually well-known to the map readers and can help them discuss results. Second, social or health data is rarely available with such a high-precision [23, 91]. Consequently, it is important to be able to display both very fine-scale population information as well as additional information at a larger scale. This is addressed in this paper through our *popchart* techniques from Sect. 6 onward.

## 3 VISUALIZING POPULATION ON CHOROPLETH MAPS

We have highlighted, based on previous work, that bivariate colour coding is not ideal for simultaneous encoding of statistical and population data. Past work has argued for the use of two separable visual variables to encode bivariate maps [32, 87]. The range of tasks or possible encoding tested has, however, been limited. We propose to test several representations combining two separable visual variables as well as visual representations combining two integrated visual variables. Based on a survey of existing techniques, we present here possible representations of bivariate choropleth maps in which the population and data of interest are simultaneously represented. For each technique, we describe previously conducted evaluations aind their outcomes.

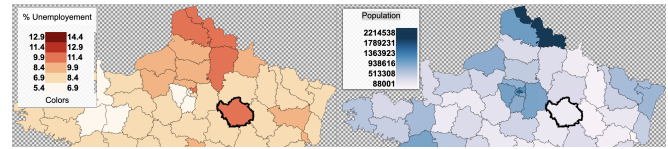

Fig. 2. Juxtaposed choropleth maps showing a statistic and population

We use maps of French 'départements' to illustrate different designs of bivariate maps. French départements are well defined administrative subdivisions of France for which population and other statistics have been recorded for a long time. We chose to use unemployment rates as statistic because it is unambiguously related to population (as a percentage) and is known to be related to population density and geography. For experimental purposes most départements have comparable sizes and so perception of colour is more homogeneous across tasks and trials. Départements are also interesting because actual inhabitants have little knowledge of their relative positions or data, and it thus decreases the possible a-priori knowledge for controlled experiments.

### 3.1 Juxtaposed Univariate Maps

Juxtaposed Univariate Maps consist of juxtaposing (i.e., putting side by side) two choropleth maps, each showing a different variable (see Fig. 2). This technique is well established and has been used, for example, to compare crime and education [4]. While Juxtaposed Univariate Maps does not fit our constraint of mapping both variables on the same figure, they are often used as a baseline for studies on bivariate choropleth maps [22]. Furthermore, mapping population directly to a second choropleth breaks Monmonier's rules [69] by showing magnitude in a choropleth map instead of intensity. For the sake of the experiment we chose to quantize the base choropleth and not the additional choropleth for population. Each map went with its corresponding legend and hovering over a region in one map also highlighted the same region in the other (see Fig. 2).

### 3.2 Absolute Maps

Absolute Maps combine the information conveyed by the two variables into a single value (Fig. 3 right). The result is likely to be very dramatic since the distribution of population is often exponential. The variation of data where, for instance, 90% of space is occupied by 10% of population is then hidden in the lowest quantile of the absolute scale. Like the population map of Juxtaposed Univariate Maps it breaks Monmonier's rule [69] by showing magnitude instead of intensity. It does, however, make sense to consider it as a baseline to the problem we are exposing in this paper.

### 3.3 colour and transparency maps

They have been introduced as value-by-alpha maps in the literature [88] as an alternative to cartograms to display enumeration data when looking at a specific measure linked to it, for example population information. Similar to Absolute Maps they can lead to a dramatic effect as 90% of space, occupied by 10% of population is hidden in the most translucent quantile (see Fig. 3 left). According to Roth's work [87], value-by-alpha maps represent asymmetrical conjunctions and are useful when one variable is more important than the other. Their use could be interesting to map population to transparency values so that low-density areas are less discernable when compared to highly populated areas.

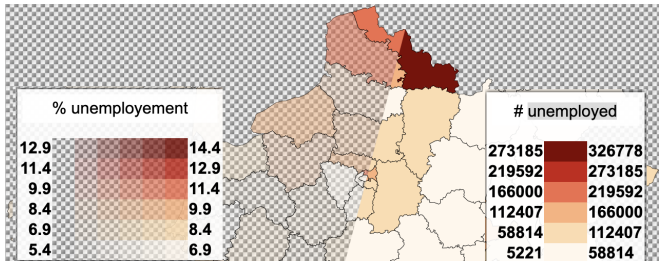
Fig. 3. Transparent map (left) and absolute choropleth (right).

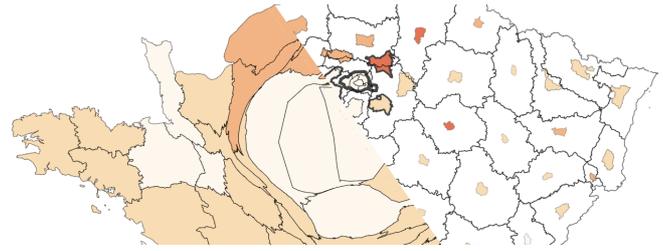
Fig. 4. Standard (left) and Non-contiguous cartograms side by side

### 3.4 3D Choropleths

The third dimension in maps is often associated with the visualization of topographic data (e.g., [18, 62]) but it can also be used to represent statistical data, as suggested in 1963 [49]. It was later applied to display the evolution of population on choropleth maps [48]. Recent technological advances on the web (e.g., WebGL) now allow easy creation of 3D representations on top of maps [66]. We therefore explore the possibility to represent population information on top of choropleth maps (see Fig. 1(C)). Unfortunately, while occlusion acts as a natural depth cue [111], 3D visualization often suffers from it as it hides data points (e.g., [33, 61, 109]). Since maps are very dense (all geographic regions are usually connected) it is also likely that regions with high-values mapped to their height will occlude regions behind them. We can however speculate, based on past research on the role of interaction for occlusion management [61, 109], that simple interaction techniques would avoid this issue. Interaction, on the other hand, can make it more difficult to see overall patterns since information from multiple views needs to be integrated mentally. We believe however that such interaction could be restricted to small rotations or pitch changes to avoid this problem while still preventing occlusion issues.

An early study [24] highlighted that 3D maps seemed to do as well as scaled-circle maps and that when faced with a 3D maps, readers are not likely to try to interpret volume but will focus on height even without a legend to guide them. This makes 3D Choroplethsa particularly interesting technique for our study. Niedomysl et al. [72] showed that printed 3D maps are less effective than printed 2D ones in recall tasks but possibly better when estimating the percentage of population living in a specific area (although the effect size was small). Stewart and Kennelly [95] showed that 3D prism maps using shadows could help in discriminating between population levels in different regions. We refrained from implementing shadows to limit the number of potential confounding factors in our experiments. In addition to their previously-highlighted advantages, 3D Choropleths maps rely on separable visual variables which are better for bivariate maps [32, 87]. We thus conjecture that this technique has the potential to effectively represent both population and the data of interest within the same map.

### 3.5 Choropleth Dot Maps

In 1865, Minard proposed to display the population in a map by adding visual elements (e.g., squares proportional to the number of inhabitants) onto each region displayed in a coloured map [68]. However, Minard never used this technique and replaced circles or squares by pie charts to depict multiple values per region. Choropleth Dot Maps also propose the display of two separable variables and is therefore better suited for a bivariate map [32, 87] and an interesting technique to evaluate.

### 3.6 Choropleth Bertillon Maps

Jacques Bertillon is mostly famous for his statistical classes but he also promoted graphical representations of statistical data [77]. His work on maps has been featured in various articles showcasing the history of statistical data visualization [36, 77]. Bertillon's work showcases maps onto which additional information is encoded by rectangles where the height and base length are dependent on two variables. In his map of the attractiveness of Paris compared with other French regions [7], Bertillon used the base of a rectangle to be proportional to the population, $P$, and the height to represent the attractiveness of Paris computed as the number of persons born in Paris divided by the number of inhabitants of the region ($I/P$) which is an index dependent on the population. Consequently, the area of the rectangle is given by $P*I/P$: the number of persons born in Paris in each region. He employed a similar technique in another map: 'Les étrangers à Paris' [8].

While area is not ideal for quantitative comparison of data [9, 20], it remains that two of the three variables presented in the map can be encoded by length which is ideal for quantitative data [9, 20]. This idea was pushed forward with, for instance, the visualization of the evolution of space taken by pasture in Normandy onto a map (see Fig M in La Statistique Graphique [58]) or by Bertillon himself in the visualization of the Operations du Mont de Piété [6]. We implemented this idea on top of a choropleth map leading to Choropleth Bertillon Maps (see Fig. 1(E)). Such representations relying on width and height are integral conjunctions [32] and are therefore unlikely to perform well for bivariate maps with different scales [32, 87]. However Choropleth Bertillon Maps also presents a redundant and separable encoding with hue and we postulate that it could yield interesting performances.

### 3.7 Deformed Cartograms

It is possible to attach the data to the regions and scale them so that their areas become proportional to their data [46]. We therefore obtain a map that uses value-by-colour for one variable and value-by-area for the population distribution (Fig. 1(D), Fig. 4 left). This idea has been exploited in statistical visualization in the 19th century [76]. An early example represented scaled-down versions of France through the years to represent the time to reach several cities with advancing transportation technology. Recent examples have used them to encode the density of population on global earthquake intensity maps [45] or residential data [43]. Cartograms, however, have inherent limitations. Their deformations impact the shape of the regions or the topology of the map, possibly hindering the readability of the map itself or the represented data [44, 88] and being able to use and interpret cartograms might require a specific learning curve [98]. However, some cartogram techniques have focused on preservation of shape or topology [78].

We are interested in contiguous cartograms [73] as they preserve neighboring regions information. While the limitations and advantages of cartograms have been thoroughly discussed [42, 78, 97] and while variations of cartograms are often compared (e.g., [73, 96, 106]), only a handful of studies compared them to other cartographic representations. Kaspar et al. [52] compared cartograms with choropleth maps with graduated circles. Their results seem to indicate that choropleth maps with graduated circles are easier to interpret but the tasks participants had to perform are not detailed in the paper. Sun and Li [96] also compared cartograms with other common representations but only asked users for subjective preferences.

### 3.8 Non Contiguous Cartogram Maps

To avoid lack of recognizability posed by standard cartograms, we tested Non Contiguous Cartogram Maps (Fig. 4 right). We shrink the size of each individual coloured region of the choropleth to represent the number of inhabitants according to the formula shrinkFactor=sqrt(density/maxdensity) where density is population/area of the considered region and maxdensity the pre-computed maximum of this variable for all regions. All regions are then zoomed in all together and only the few biggest ones are translated to avoid overlapping. The regions with the highest number of inhabitants are represented bigger than their correct geographical size while all other regions will necessarily shrink. This trade-off between very few regions scaled up and translated and the vast majority of regions scaled down and correctly centered ensures good readability of topology and choropleth colours. The blank spaces between the regions allow the correct frontiers to be displayed between them and can help reading individual colours without influence from the neighbours. Yet, it may also degrade the viewer's ability to compare colours between two adjacent regions. This problem is softened by the choice of quantized choropleth maps which produce very distinct colours. We expect Non Contiguous Cartogram Maps to be easier to interpret because, unlike Deformed Cartograms, they preserves the original shape of the regions and the topology.

## 4 INITIAL EXPERIMENTATION AND EXPERT FEEDBACK

### 4.1 Experiment description

To select a subset of techniques for a controlled experiment, we ran an initial online experiment to collect feedback from visualization researchers (from several institutions) about the techniques described in Sect. 3. We showed screenshots of each visualization, displaying unemployment and population distribution in French départments. We asked them to give benefits and limitations of each representation and to rank, on a Likert-scale from 1 (Very Easy) to 5 (Very Hard), how easy it was for them to achieve specific tasks with each visual representation.

### 4.2 Tasks

To explore the benefits and limitations of each representation, we defined a series of specific tasks based on a review of cartography tasks. Roth [86] classified high-level user goals, objectives, operand primitives and operators for tasks in cartography. Of particular importance for our study is the categorization of operands (Space-Alone, Attributes-in-Space, or Space-in-Time) and objectives (Identify, Compare, Rank, Associate and Delineate). While comprehensive, this classification is aimed at monovariate maps and so most of the tasks and objectives described are less appropriate in the testing of bivariate maps. Furthermore, the taxonomy does not directly describe a specific objective that is often considered important in the case of cartograms: *summarize*. This summarization task refers to the objective when users are tempted, or even asked, to aggregate data in some parts of the map to try to form and memorize a "big picture" message. It is one of the elements of a cartograms' taxonomy of goals [74],frequently used to compare different types of cartograms (e.g., [73]). This summarize objective is often presented in information visualization taxonomies by the names *summarize* (e.g., [16, 112]), *overview* (e.g., [57, 92]), or even *review* [93]. While not initially used to define tasks in bivariate maps, Roth's taxonomy [86] can, nonetheless, be used to describe tasks with normalized and widely-accepted vocabulary. In Table 1 we describe the set of 5 questions that we have derived using this taxonomy.

### 4.3 Results

Ten visualization researchers answered this online experiment. Results of the Likert-scale ratings are presented in Fig. 5. We counted the number of times when the number of (Very) Difficult is higher or equal to the number of (Very) Easy for each technique and each question. If it happened more than twice for a technique, we looked at the qualitative feedback to determine whether we should exclude them from the user study. Non Contiguous Cartogram Maps ($Q1_{Pop}$, $Q3_{Comb}$, $Q5_{Sum}$), Deformed Cartograms ($Q3_{Comb}$, $Q4_{Comb}$, $Q5_{Sum}$), Transparent maps ($Q1_{Pop}$, $Q2_{Ump}$, $Q3_{Comb}$, $Q4_{Comb}$), and Juxtaposed Univariate Maps ($Q3_{Comb}$, $Q4_{Comb}$, $Q5_{Sum}$) were candidates to be excluded.

The feedback for Transparent maps confirmed that participants had issues understanding the mapping (×8 participants), that it was difficult to know whether the colour or the transparency was changing (×4), and that colours were hard to distinguish at lower levels of transparency (for low-populated areas). We thus removed it from the pool of techniques for the controlled experiment. Similarly, while Non Contiguous Cartogram Maps were generally praised for the concept, 6 participants reported that regions were usually too small to see properly/compare. We therefore also excluded it from the pool of techniques.

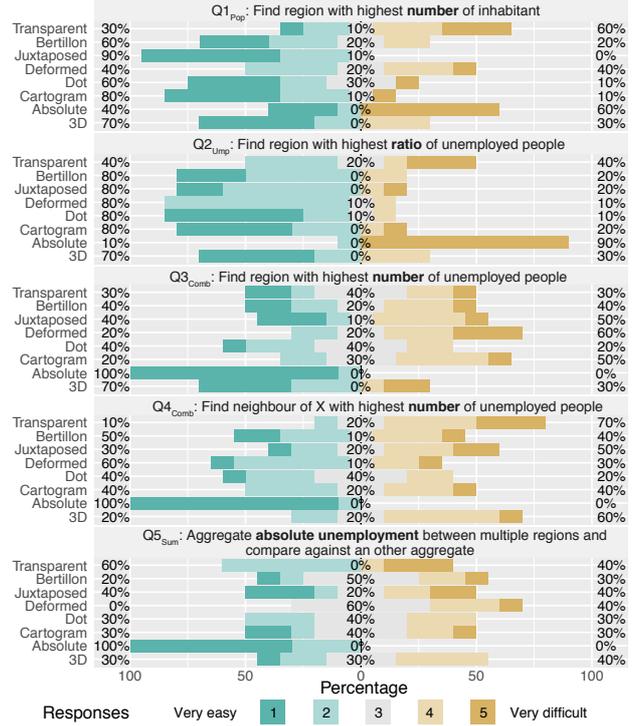

Fig. 5. Answers to Likert-scale ratings for the first level of granularity.

Juxtaposed Univariate Maps were not described very negatively in the feedback, apart from the observation that the computation of the absolute number (results of the two juxtaposed maps) could be difficult. Since it is often used and the baseline of some other studies (e.g., [22]), we decided to retain to allow comparison with other techniques. Similarly, while Deformed Cartograms were not generally well perceived (6 participants reported that they were hard to interpret, and 3 said that it was difficult to compare regions), they are popular and heavily studied (e.g., [1, 64]), so we kept them in our experiment.

Based on qualitative feedback only, we removed two other representations. First, Absolute Maps were, as expected, regarded as missing essential information to complete the tasks, so we removed them. Then, Choropleth Dot Maps are very similar to the idea of the Choropleth Bertillon Maps and we therefore decided to remove them from the pool of techniques. Finally, 7 participants complained about the label placement, mirroring past findings (e.g., [63, 89]). In our controlled study, we thus remove labels and only display them when hovering.

We thus established a list of 4 techniques to use in a controlled experiment: Juxtaposed Univariate Maps, Choropleth Bertillon Maps, 3D Choropleths, and Deformed Cartograms.

## 5 CONTROLLED EXPERIMENT 1

We conducted a controlled experiment to evaluate which of the retained techniques would be helpful to show the distribution of population and the statistic of interest. The experiment is available at tiny.cc/mapstudy and its pre-registration (frozen data-analysis scripts) at osf.io/rgkpj.

### 5.1 Participants, Questions, and Techniques

Invitations to participate, with a link to the study, were sent by email to students and researchers working both within and outside of the fields of visualization or computer science. The email also asked them to forward the experiment to others. As such, predicting the number of respondents was difficult and we, instead, decided to stop the data collection on a specific date (after 10 days). While not common in HCI, using a time-based stopping rule to preregister a sample size is not rare and is actually in the template of preregistrations on AsPredicted.org. The techniques we used in the experiment were the ones that we had previously identified (namely, Juxtaposed Univariate Maps, Choropleth

Table 1. Description and naming of the tasks we used in our experiment

| Name | Description | Break down according to Roth's taxonomy [86] |
|------|-------------|---------------------------------------------|
| $Q1_{Pop}$ | Read the number of inhabitants in a specific region. | RANK attributes in space for population |
| $Q2_{Ump}$ | Read the statistic of interest (unemployment) in a region. | RANK in attributes in space for unemployment |
| $Q3_{Comb}$ | Combine population and unemployment to find the region with the highest number of unemployed people. | RANK in attributes in space for unemployment × population |
| $Q4_{Comb}$ | Identify the neighbour of a region that has the highest number of unemployed people. | RANK in attributes in space for unemployment × population / IDENTIFY in space alone |
| $Q5_{Sum}$ | Average the absolute number of unemployed people over multiple regions and compare with another average | SUMMARIZE (from [74]) for unemployment × population / RANK for both summarized regions |

Bertillon Maps, 3D Choropleths, and Deformed Cartograms), and we again used the data of France's unemployment and population. The order of techniques was counterbalanced to avoid learning effects.

The set of questions was taken from our previous experiment. To avoid learning effects, 4 different but equivalent sets of questions were prepared. Each set contained 5 questions as previously described ($Q1_{Pop}$ to $Q5_{Sum}$) and were assigned, in turn, to different representation: the same set was not always asked when using the same visualization. For each question, areas of interest were highlighted so that the completion time did not contain the search time but only the time it took for participants to answer a question with a given representation.

We did not give users the possibility to interact (rotations, dragging, zooming, or changing the pitch). While this does not reflect the real usage of such maps, it is necessary to avoid possible confounding factors in our experiments, and is not uncommon in visualization experiments [14, 26, 103]. While this could negatively affect 3D Choropleths, it is essential to have a fair comparison between techniques.

### 5.2 Planned Analysis Results

A total of 60 participants completed our experiment. However, 2 entered nonsensical demographics and were removed according to our preregistered exclusion criteria. We thus had valid data from 58 participants (19 females, mean age = 26.3, median = 26, SD = 4.8, range 18-41). While such data is usually analysed with NHST and ANOVAs, recent criticism of NHST to analyze experimental data [3, 25, 29, 38, 40, 65] and recent APA recommendations [101], led us to report our results using estimation techniques with effect sizes[1] and confidence intervals instead of $p$-values.[2] We interpret them as providing different strengths of evidence about the population mean [11, 12, 25, 28, 39].

#### 5.2.1 Accuracy

Accuracy results, whether a participant had the right answer (score of 1) or a wrong answer (score of 0), are presented in Fig. 6, while Fig. 7 presents pair-wise differences (i.e., individual differences). The small overlap of confidence intervals (CIs) for $Q1_{Pop}$ suggests that Juxtaposed Univariate Maps is likely to perform better than the other techniques. This is confirmed by the non-overlap with 0 in Fig. 7. Similarly, for $Q2_{Ump}$, the small overlap of CIs suggests that Juxtaposed Univariate Maps and Deformed Cartograms lead to a better accuracy (confirmed in Fig. 7). Concerning $Q3_{Comb}$, we could not find evidence of a difference between the techniques which all seemed to perform poorly (around 50% accuracy). The small overlap of CIs in Fig. 6 provides evidence that Juxtaposed Univariate Maps and Choropleth Bertillon Maps can outperform the other two techniques for $Q4_{Comb}$ (confirmed in Fig. 7). Finally, for $Q5_{Sum}$ the CIs in both figures provide weak evidence that 3D Choropleths and Deformed Cartograms outperform the other two techniques. In all cases where evidence of a difference are observed, the difference ranges from 13% to 22%.

---
[1]Effect size refers to the means we measured. We do not use standardized effect sizes [21]: reporting them is not always recommended [2].
[2]A $p$-value-approach reading of our results can still be inferred [54].

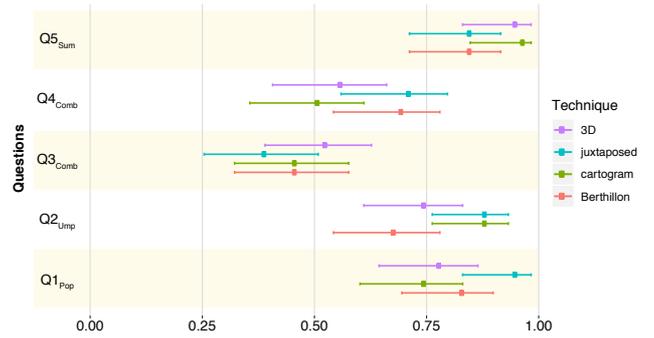

Fig. 6. Mean accuracy. Error bars: 95% Bootstrap Confidence Intervals.

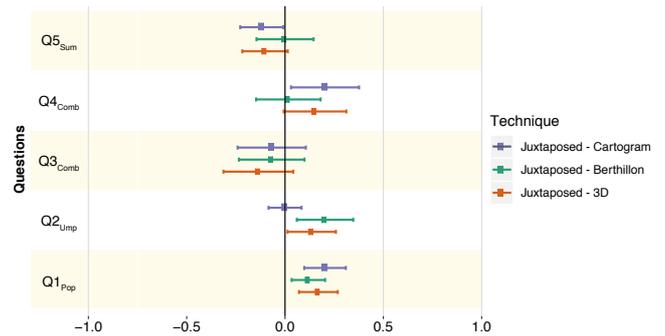

Fig. 7. Pairwise differences in accuracy. Error bars: 95% Bootstrap CIs.

#### 5.2.2 Completion Time

We analyzed log-transformed data to correct for positive skewedness and present antilogged results as is standard for such data analysis processes [90] and common in HCI (e.g., [10, 13, 47, 56, 110]). Consequently, we arrive at geometric means.[3] They dampen the effect of extreme trial completion times which tend to bias an arithmetic mean. Results and pair-wise ratios are plotted respectively in Fig. 8 and Fig. 9.

For $Q1_{Pop}$, the CIs in Fig. 8 give evidence that 3D Choropleths can be faster than the other techniques. Fig. 9 confirm this finding and suggest that Choropleth Bertillon Maps could be slower, even though the effect is smaller. For $Q2_{Ump}$, our results seem to suggest that Choropleth Bertillon Maps are slower than the other equally fast techniques. This is confirmed by the CIs in Fig. 9, providing strong evidence that Juxtaposed Univariate Maps (and consequently the other two techniques too) are almost twice as fast as Choropleth Bertillon Maps. The small overlap between Juxtaposed Univariate Maps and Deformed Cartograms in Fig. 8 seem to suggest than Juxtaposed Univariate Maps could be slower than Deformed Cartograms for $Q3_{Comb}$. In addition, Fig. 9 provides strong evidence of that difference: Juxtaposed Univariate Maps is approximately 1.3 times slower than Deformed Cartograms. For both $Q4_{Comb}$ and $Q5_{Sum}$, the small overlap between Juxtaposed Univariate Maps and the other techniques provides evidence of a difference which is confirmed in Fig. 9: Juxtaposed Univariate Maps is approximately 1.3 times slower than the other three techniques for $Q4_{Comb}$ and approximately 1.4 times slower than the other three techniques for $Q5_{Sum}$.

---
[3]An arithmetic mean uses the sum of values, a geometric mean uses the product of values.

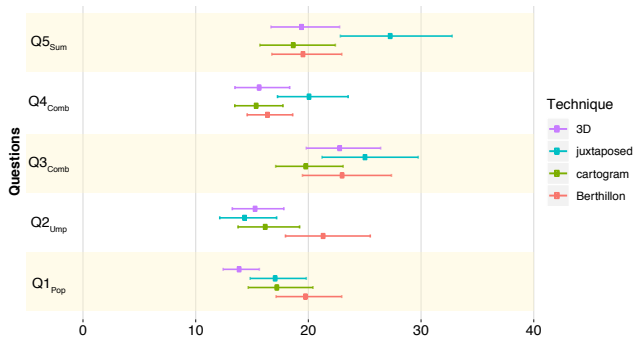

Fig. 8. Mean completion time (lower is better). Error bars: 95% CIs.

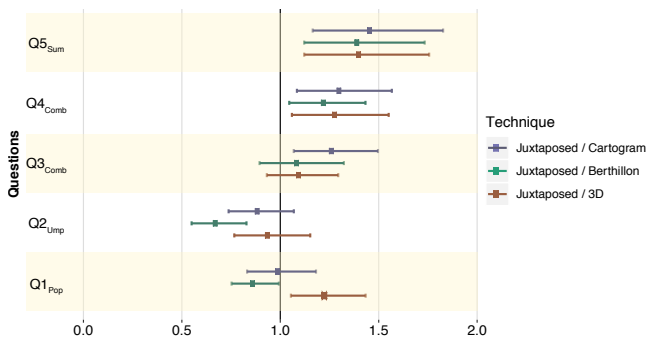

Fig. 9. Pairwise ratios for completion time. Error bars: 95% CIs.

### 5.2.3 Ranking

Participants' ranking is shown in Table 2, as planned in the pre-registration. We see that 3D Choropleths was ranked as the favorite technique by the most participants with Choropleth Bertillon Maps not far behind. Deformed Cartograms was ranked as the least favorite by the highest number of participants (23) with the worst mean and median, indicating that this technique is the least favorite overall. The best median scores are obtained by Juxtaposed Univariate Maps, Choropleth Bertillon Maps, and 3D Choropleths while the best mean is obtained by 3D Choropleths. This highlights that there is not a clear favorite technique, but that there might be a slight preference for 3D Choropleths.

### 5.3 Additional analysis: qualitative feedback

Participants could also comment on the limitations and benefits of each technique. The fully categorized and raw data is available on osf.io/8rxwg/. We summarize here the main insights. Juxtaposed Univariate Maps were reported to be a standard and simple visualization ($\times$ 5 participants) and to provide readable information thanks to the two separated maps ($\times$ 12). On the other hand, they were said to make the combination of both variables hard ($\times$ 15). Having to switch between two maps was also reported as a drawback ($\times$ 6). For 3D Choropleths, the main reported issues were: occlusion ($\times$ 23), and the difficulty to compare widely separated regions ($\times$ 4). Concerning their benefits, it was reported that interaction would make them better—and solve the occlusion problem— ($\times$ 5), that comparisons are easy to make ($\times$ 5), that the maps are visually appealing ($\times$ 5), and that it was easy to combine both pieces of information provided by the visual variables ($\times$ 12), though the opposite was also reported ($\times$ 2). Deformed Cartograms were reported to make it hard to compare population information (difficulty to compare differently-shaped regions, $\times$ 19); to destroy the geography of the country ($\times$ 11), to be ugly ($\times$ 2), and to make sparsely populated regions hard to see ($\times$ 2). Nonetheless, they were reported to be clear ($\times$ 2) and visually appealing or interesting ($\times$ 4). Finally, Choropleth Bertillon Maps were reported to make all the information easily accessible without having to mentally combine them ($\times$ 15), but also to create a lot of overlap ($\times$ 16), to be not visually pleasing, or even to be ugly ($\times$ 5), and to make region comparisons difficult ($\times$ 5).

Table 2. Ranking between most (1) and least (4) favorite technique.

| Technique | Mean | Med | SD | 1st | 2nd | 3rd | 4th |
|---|---|---|---|---|---|---|---|
| Juxtaposed Univariate | 2.4 | 2 | 0.9 | 10 | 19 | 19 | 1 |
| 3D Choropleth Maps | 2.2 | 2 | 1.1 | 19 | 12 | 14 | 9 |
| Deformed Cartogram | 2.8 | 3 | 1.1 | 8 | 14 | 9 | 23 |
| Choropleth Bertillon | 2.3 | 2 | 1.2 | 17 | 15 | 9 | 13 |

### 5.4 Discussions

Unsurprisingly, when focusing on a single variable ($Q1_{Pop}$ and $Q2_{Ump}$), Juxtaposed Univariate Maps provide a higher accuracy than other techniques and are relatively fast. When combining the two variables ($Q3_{Comb}$) it seems that most techniques would perform equally well in completion time and accuracy. This reinforces previous findings comparing some of the designs that we have tested [32]. For $Q4_{Comb}$, asking participants to combine both variables and to consider geographical information, Juxtaposed Univariate Maps and Choropleth Bertillon Maps seem to give a better accuracy although Juxtaposed Univariate Maps maps are much slower. Finally, 3D Choropleths and Deformed Cartograms seem to lead to particularly accurate answers when participants are asked to combine both variables and aggregate this over several regions ($Q5_{Sum}$) and are also much faster than Juxtaposed Univariate Maps. These results align with previous work suggesting that Deformed Cartograms can be useful for *summarizing* tasks. However, 3D Choropleths have not been suggested as interesting *summarizing* techniques, and our findings are thus more surprising.

Overall, Deformed Cartograms were not really appreciated by participants and only proved to be better for *summarizing*. Our qualitative results also mirror past results: the deformations hinder the understanding of the map [43, 88]. 3D Choropleths were very appreciated and could be easily improved with interaction to avoid occlusion issues. They provide the information very quickly (Fig. 8) but are not the most accurate technique. The accuracy they have for *summarize* tasks make them an excellent candidate to replace Deformed Cartograms.

Juxtaposed Univariate Maps, often considered as a baseline, are among the most accurate techniques for tasks relying on *locating* or *comparing* (mirroring previous findings e.g., [22]), and could thus still be used as a baseline in future studies. Their accuracy, however, seems to come at the cost of time. Finally, despite their popularity and the positive feedback, Choropleth Bertillon Maps did not outperform Juxtaposed Univariate Maps but always had a good overall accuracy, showing that glyphs on choropleth maps can exhibit good performances.

Our data suggest that the accuracy was poor when participants combined information from the two maps or visual variables (~50% for $Q3_{Comb}$, ~65% for $Q4_{Comb}$). Currently most choropleth maps displaying social data do not include the repartition of the population (e.g., [53, 107]) which can be dramatically misleading [45, 78, 88]. However, even if they showed the distribution of population, our results indicate that such maps could still be misunderstood. While past work [32, 86] suggests that the techniques we tested should perform well (because they are made of two separable visual variables) our results indicate that they do not give high accuracy. This suggests that new techniques should be developed and investigated for this type of task. Since the task in $Q3_{Comb}$ was quite similar to that in $Q4_{Comb}$, we can hypothesize that combining information from two visual variables on a map requires some training before being accurate enough.

Finally, despite our quite large sample size, we observe in the figures

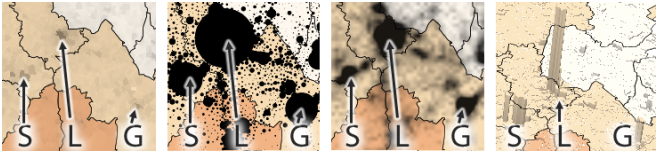

Fig. 10. Cities of St Etienne(S), Lyon(L) and Grenoble(G) at higher level of detail using Popchart dasymetric overlay maps, Popchart Dot Maps, Popchart Heatmap Maps and Popchart 3D PrismMaps

focusing on accuracy that the confidence intervals are still quite large for all questions. This suggests that the best visual representation to increase accuracy of interpretation could depend on the individual user.

## 6 POPCHART: FINE-SCALE POPULATION ON CHOROPLETHS

To exploit the potential benefits of dasymetric maps, as described earlier, we have developed popchart which aims at overlaying data on a choropleth map to augment the fine-scaled population distribution it shows. We have explored four representations described in the following subsections.

### 6.1 Popchart dasymetric overlay maps

Popchart dasymetric overlay maps overlay a transparent choropleth (also called dasymetric map) on top of the standard choropleth to convey population. With two levels of granularity we use another transparent colour on top of the coloured region to display population. In all the maps we used black to encode local population data. (Fig. 10a) shows large cities and suburbs but the larger population density along valleys and rivers are not clearly visible. Here the total population is given by the sum of the perceived opacities. This technique can suffer from the same perception bias as choropleth maps in that a city presenting a larger area can be perceived as having a greater total population.

### 6.2 Popchart Dot Maps

Popchart Dot Maps use dots to represent population per city. Large dots can overlap suburbs as seen in Fig. 10b for Lyon and St Etienne. Still, it clearly shows population distribution along rivers, coasts and valleys (e.g., south of Lyon). Here the total population is given by the sum of the sizes of the circles (when they are visible).

### 6.3 Popchart Heatmap Maps

Popchart Heatmap Maps use heatmap overlays. Heatmap is a popular technique to aggregate a large number of points to display on a map. We use the default package of our Geo-Information System to produce a heatmap from transparent (low population density) to black (high population density). Large cities are visible but their centre can be harder to identify: the 'fuzzy' nature of heatmaps makes it difficult to distinguish the distribution of population between a main city and its suburbs. For instance, in Fig. 10c, it is difficult to see that there are more people in the East of Lyon than in the West. Here the total population is relative to the total amount of 'ink' present in a region.

### 6.4 Popchart 3D PrismMaps

Popchart 3D PrismMaps, relies on height to display the number of inhabitants at the city level. The 3D extrusion for each city is based on the limits of the city and the height is defined by the number of inhabitants. The colour of the top and border of the 3D volumes encodes the statistical value associated with the enclosing region. As people tend to be concentrated in a few big cities, most of the volumes are very flat making the resulting map appear similar to a tilted standard choropleth. Among these flat coloured volumes one can distinguish big cities as very high 'skyscrapers' (Fig. 10d) sometimes surrounded by lower 'towers' in the suburbs. Only a few cities, behind the biggest ones, tend to be occluded by the 3D perspective. The total population of a given region is given by the sum of all the heights of the volumes.

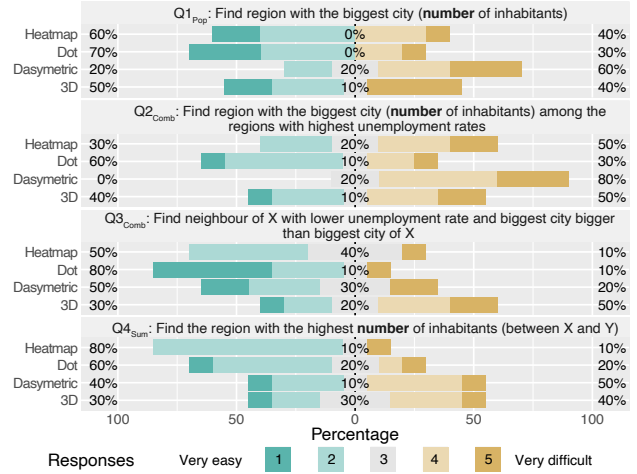

Fig. 11. Answers to Likert-scale ratings from visualization researchers.

## 7 INITIAL EXPERIMENT WITH VISUALIZATION RESEARCHERS

To find a set of techniques for a controlled experiment, we conducted an investigation with visualization researchers to get feedback on the techniques described in Sect. 6. We follow the same procedure described in Sect. 4 and describe our tasks (Table 3) using Roth's taxonomy [86].

The same 10 visualization researchers answered. Results are presented in Fig. 11. We counted how often the number of (Very) Difficult is greater or equal to the number of (Very) Easy scores for each technique. If it happened more than twice, we looked at the qualitative feedback to determine whether we should exclude a technique. Popchart dasymetric overlay maps ($Q2_{Comb}$, $Q3_{Comb}$, $Q4_{Sum}$) and Popchart 3D PrismMaps ($Q2_{Comb}$, $Q3_{Comb}$, $Q4_{Sum}$) were candidates to be excluded.

Popchart dasymetric overlay maps was reported as difficult to compare ($\times$ 3) and interpret ($\times$ 4). We therefore excluded it from our experiment. Popchart 3D PrismMaps was, despite its low ratings, generally well perceived (4 participants reported that all information could be computed with this technique) so we kept it in the pool of techniques.

Based on this experiment, we thus decided to keep popchart 3D prismMaps, popchart Heatmap maps, and Popchart Dot Maps.

## 8 CONTROLLED EXPERIMENT 2

We conducted a second experiment to evaluate which of the previously defined techniques would be more helpful to convey the spatial distribution and the statistic of interest with a fine-scale distribution of population. The code used for the experiment is available at tiny.cc/mapstudy along with a frozen preregistration at osf.io/svh8a.

### 8.1 Participants, Questions, and Techniques

Invitations to participate were sent by email, including a link to the study, to students and researchers working in computer science and in other fields. The email also invited them to forward the experiment.

We again used a time-based stopping rule (10days) to preregister a sample size. The previously identified techniques used in the experiment are popchart 3D prismMaps, popchart Heatmap maps, Popchart Dot Maps in a counterbalanced order to avoid learning effects. We also used the data on France's unemployment and population.

The tasks were also taken from the initial experiment with researchers. To avoid learning effects, 3 different, but equivalent, sets of questions were prepared. Each set contained 4 questions ($Q1_{Pop}$ to $Q4_{Sum}$) and were assigned, in turn, to different visualization techniques so that the same set of questions was not always asked about the same visualization technique. For each question, areas of interest were highlighted so that the completion time does not contain the search time but only the time it took participants to answer the question. To limit possible confounding factors we did not give participants the ability to interact with the map.

Table 3. Description and naming of the tasks we used in our experiment

| Name | Description | Break down according to Roth's taxonomy [86] |
|---|---|---|
| $Q1_{Pop}$ | Find which region has the biggest city, in terms of population. | RANK attributes in space for population (city level) |
| $Q2_{Comb}$ | Identify the region with the lowest/highest unemployment rate in a specific area and then determine which one has the city with the highest number of inhabitants. | RANK in attributes in space for unemployment (region level)<br>RANK in attributes in space for population (city level) |
| $Q3_{Comb}$ | Compare the unemployment rate of a region with its neighbours (i.e, find neighbours with higher/lower/similar statistical value) and find which neighbour has the biggest city. | RANK in attributes in space for unemployment (region level)<br>RANK in attributes in space for population (city level)<br>IDENTIFY in space alone |
| $Q4_{Sum}$ | Average the population information presented at the city level to a region level: determining whether one region has a higher number of inhabitants than another. | SUMMARIZE (from [74]) for population (city level to region level)<br>RANK for both summarized regions |

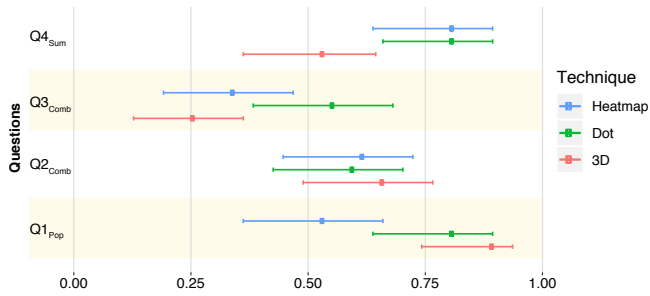

Fig. 12. Mean accuracy (higher = better). Error bars: 95% Bootstrap CIs.

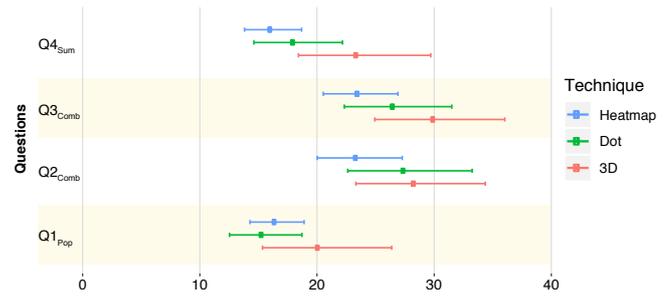

Fig. 14. Mean completion time (lower = better). Error bars: 95% CIs.

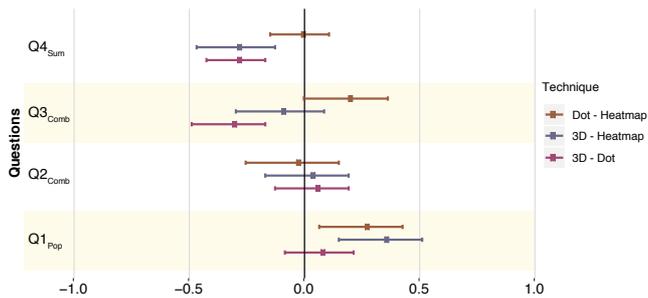

Fig. 13. Pairwise differences in accuracy. Error bars: 95% Bootstrap CIs.

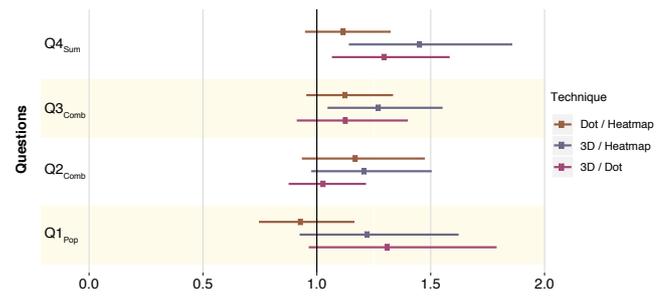

Fig. 15. Pairwise ratios in time. Error bars: 95% CIs.

## 8.2 Planned Analysis Results

A total of 47 participants (17 females, aged from 19 to 57, mean = 26.8, median = 26, SD = 6.8) completed this experiment. None of their data was excluded with our preregistered exclusion criteria. Data analysis is conducted following the methodology described in Sect. 5.2.

### 8.2.1 Accuracy

Accuracy results are shown in Fig. 12, pairwise differences in Fig. 13. The small overlap between Popchart Heatmap Maps and the other two techniques in Fig. 12 shows that Popchart Heatmap Maps has a lower accuracy than the other two, performing similarly well (confirmed in Fig. 13), for $Q1_{Pop}$. Concerning $Q2_{Comb}$ the CIs in both figures indicate no difference between the 3 techniques. For $Q3_{Comb}$ the configuration of the CIs strongly indicates that Popchart Dot Maps is more accurate than Popchart 3D PrismMaps and we see evidence that Popchart Dot Maps is more accurate than Popchart Heatmap Maps with a smaller effect size. Finally, for $Q4_{Sum}$ we see no evidence for a difference between Popchart Dot Maps and Popchart Heatmap Maps but both are more accurate than Popchart 3D PrismMaps. This is confirmed in Fig. 13.

### 8.2.2 Completion Time

The results are presented in Fig. 14 and pairwise ratios in Fig. 15. For $Q1_{Pop}$, while the overlap of CIs would suggest that Popchart 3D PrismMaps is slower than the other two techniques, individual differences in Fig. 15 leads us to argue that evidence is quite weak in this case. The configuration of CIs in both figures for $Q2_{Comb}$ and $Q3_{Comb}$ suggest that our data does not provide evidence of a difference between the 3 techniques. For $Q4_{Sum}$, we have strong evidence that Popchart 3D PrismMaps is almost 1.5 times slower than Popchart Heatmap Maps and approximately 1.3 times slower than popchart dot mapswhile these two techniques seem to achieve similar performance.

### 8.2.3 Ranking

The ranking data is presented in Table 4. We can see that Popchart Dot Maps were most frequently ranked as the favorite technique. Popchart Heatmap Maps and Popchart 3D PrismMaps have similar results in this respect, but Popchart 3D PrismMaps were ranked as the least favorite by the highest number of participants ($\times$ 19) and has the worst mean. This suggests that this technique could be the least popular overall. The best mean score was obtained by Popchart Dot Maps which seems to be the overall favorite. All differences are, however, quite small.

## 8.3 Additional analysis: qualitative feedback

Participants were invited to reflect on the limitations and benefits of the techniques. The fully categorized and raw data is available on osf.io/8rxwg/. Popchart 3D PrismMaps were praised for their capacity to increase readability ($\times$ 2 participants), to avoid overlaying information on top of the map ($\times$ 3). However, the information was considered hard to aggregate over a region ($\times$ 3), the angle of the map was reported as problematic ($\times$ 6), smaller cities were difficult to perceive ($\times$ 7), the 3D bars were too transparent ($\times$ 5) and height was said to be hard to compare ($\times$ 7). Popchart Dot Maps were reported as easy to read ($\times$ 13), a standard visualization ($\times$ 2) even though the circles representing cities could occlude other cities ($\times$ 18) or occlude the geographical information or even the information presented by the choropleth itself ($\times$ 18). Popchart Heatmap Maps were reported as easy to read ($\times$ 9) but participants stated that they were hard to compare ($\times$ 6), could hide geographical information ($\times$ 7) and were not visually pleasing ($\times$ 3).

Table 4. Ranking between most (1) and least (3) favorite technique

| Technique | Mean | Median | SD | #1st | #2nd | #3rd |
|---|---|---|---|---|---|---|
| Popchart 3D PrismMaps | 2.2 | 2 | 0.8 | 11 | 12 | 19 |
| Popchart Dot Maps | 1.8 | 2 | 0.8 | 18 | 14 | 10 |
| Popchart Heatmap Maps | 2 | 2 | 0.8 | 12 | 18 | 12 |

### 8.4 Discussion

To identify the city with the highest population density, Popchart 3D PrismMaps seem to show comparable accuracy (but worse completion time) as other methods except for $Q3_{Comb}$ asking for geographical information to be considered. There Popchart 3D PrismMaps performed poorly. This could be explained by the extra geographical task.

For combination of unemployment data and population of some cities ($Q2_{Comb}$ and $Q3_{Comb}$) it seems that Popchart Dot Maps might have a slight advantage in terms of accuracy—visible mostly in $Q3_{Comb}$. All techniques seemed to exhibit a similar completion time.

To average the population of an entire region ($Q4_{Sum}$), Popchart 3D PrismMaps performed the worst in terms of both time and accuracy, while the other two techniques seemed to perform equally well. Interestingly, Popchart Dot Maps and Popchart Heatmap Maps perform very well for such tasks (~80%), thus validating that one can display information at a finer scale while still leaving the possibility for users to aggregate per region quite accurately. We can deduce that readers of Popchart maps can still answer most of the tasks that were presented in the first experiment we conducted. This is a strong advantage over the techniques explained and studied in Sect. 3: Popchart maps can therefore provide even more precise information without sacrificing the original information conveyed by choropleth maps.

The low performance for tasks focusing on a city population and unemployment of a region ($Q2_{Comb}$ and $Q3_{Comb}$ with ~65% accuracy and, at most, 55% accuracy) is also particularly interesting. It calls for more techniques to tackle the issue. This is supported by the qualitative feedback gathered which highlights that the technique that performs the worst (Popchart 3D PrismMaps) has the advantage over the other two of not hiding any kind of information on the map, as was identified as a drawback of both Popchart Dot Maps and Popchart Heatmap Maps.

## 9 Conclusion, Limitations, and Future Work

With the development of web technologies, maps can now easily be made and shared by lay people or researchers in social sciences [66]. In this context, understanding the pitfalls and benefits of specific designs for communication of research results to lay people is therefore particularly important, in particular since choropleth maps are inherently prone to lead to misinterpretation [1, 44, 64, 88, 94]. With that objective in mind we have proposed several visual representations of bivariate choropleth maps and evaluated them through two studies.

Our experiments of course have some limitations. First of all we could have considered additional techniques such as colour-based bivariate maps. Maps have been studied quite extensively and it was impossible to be exhaustive. Another limitation lies in the loading time of the maps which influenced the completion time. To avoid this issue we could have used simple screenshots but we would have lost all interactivity (e.g., providing labels on hovering) and opting for such a solution would have made the experiment much harder. On a similar note, interaction could have changed the performance of some techniques (notably for techniques relying on 3D graphics). This is something we intend to investigate in future work. While giving the possibility to rotate/drag the map and change the viewing angle would help overcome most of the limitations of maps relying on 3D graphics (in our case popchart 3D prismMaps) they could also lead to problems in interpretation. Indeed, arbitrarily-rotated maps could be more difficult to make sense of if the user cannot recognize the geography. Giving the possibility for participants to interact may also have led to longer completion times, or a much higher cognitive load if such interactions are carried out through classical input techniques [13]. We hypothesize, however, that constrained interaction allowing only small adjustments of the pitch/viewing angle and small rotations can lead to ideal performance by overcoming the occlusion issue while keeping the map close to a specific orientation that does not impede the user's sense-making process and would limit the impact on cognitive load. Finally, we have currently tested our technique with a single dataset and studies with different datasets should further confirm our results.

While 3D representations are seldom used in information visualization, our study results are consistent with previous studies and highlight the potential of 3D Choropleths, performing as well as deformed cartogram maps in particular for tasks involving *Summarizing* information from different regions to get a big-picture message, for which deformed cartogram maps have been praised [74]. For both dasymetric and regular choropleth maps, the results we have obtained are promising but the interpretation of population distribution combined with a statistic of interest remains a difficult problem with all the investigated representations. These results strongly highlight the need for more work on bivariate choropleth maps displaying population distribution. Our work paves the way towards a better understanding of the interpretation of bivariate choropleth maps. In addition, the qualitative feedback we have obtained also opens the way to intriguing possible further developments of our techniques. For example, 130 years after Bertillon's work, our studies raise the possibility to use and redesign (possibly with a 3rd dimension) Bertillon's rectangles to strengthen the perception of the product of the two variables. The identified advantages of popchart 3D prismMaps over the other visual representations considered in this experiment could also suggest that Popchart 3D PrismMaps might be significantly improved with only a small amount of redesign or with the addition of interaction techniques. Finally, we would like to investigate what interaction technique should be supported for maps combining differently-scaled variables and what tasks they should support.

### Acknowledgments

We thank Jean-Daniel Fekete for his feedback and our participants.

### References

[1] M. J. Alam, S. G. Kobourov, and S. Veeramoni. Quantitative measures for cartogram generation techniques. *Computer Graphics Forum*, 34(3):351–360, 2015. doi: 10.1111/cgf.12647

[2] T. Baguley. Standardized or simple effect size: What should be reported? *British Journal of Psychology*, 100(3):603–617, Aug. 2009. doi: 10.1348/000712608X377117

[3] M. Baker. Statisticians issue warning over misuse of P values. *Nature*, 531(7593):151, Mar. 2016. doi: 10.1038/nature.2016.19503

[4] A. Balbi and A.-M. Guerry. Statistique comparée de l'état de l'instruction et du nombre des crimes, 1829.

[5] L. V. Barrozo, R. P. Pérez-Machado, C. Small, and W. Cabral-Miranda. Changing spatial perception: dasymetric mapping to improve analysis of health outcomes in a megacity. *Journal of Maps*, 12(5):1242–1247, 2016. doi: 10.1080/17445647.2015.1101403

[6] J. Bertillon. *Atlas de statistique graphique de la ville de Paris, année 1889, tome II*. Préfecture du département de la Seine, Secrétariat général, Service de la statistique municipale., 1891.

[7] J. Bertillon. *Résultats statistiques du dénombrement de 1891 pour la ville de Paris et le département de la Seine et renseignements relatifs aux dénombrements antérieurs*, p. 73. Préfecture de la Seine, 1891.

[8] J. Bertillon. *Résultats statistiques du dénombrement de 1891 pour la ville de Paris et le département de la Seine et renseignements relatifs aux dénombrements antérieurs*, p. 84. Préfecture de la Seine, 1891.

[9] J. Bertin. *Sémiologie graphique: les diagrammes*. Mouton, 1967.

[10] L. Besançon, M. Ammi, and T. Isenberg. Pressure-Based Gain Factor Control for Mobile 3D Interaction using Locally-Coupled Devices. In

*Proc. CHI*, pp. 1831–1842. Denver, May 2017. doi: 10.1145/3025453.3025890

[11] L. Besançon and P. Dragicevic. La différence significative entre valeurs p et intervalles de confiance (The significant difference between p-values and confidence intervals). In *IHM*, 2017. https://hal.inria.fr/hal-01562281.

[12] L. Besançon and P. Dragicevic. The Continued Prevalence of Dichotomous Inferences at CHI. In *CHI Extended Abstracts*. Glasgow, United Kingdom, May 2019. doi: 10.1145/3290607.3310432

[13] L. Besançon, P. Issartel, M. Ammi, and T. Isenberg. Mouse, tactile, and tangible input for 3D manipulation. In *Proc. CHI*, pp. 4727–4740. ACM, New York, 2017. doi: 10.1145/3025453.3025863

[14] T. Blascheck, L. Besançon, A. Bezerianos, B. Lee, and P. Isenberg. Glanceable visualization: Studies of data comparison performance on smartwatches. *IEEE TVCG*, 25, January 2019. doi: 10.1109/TVCG.2018.2865142

[15] D. Borland and R. M. Taylor Ii. Rainbow color map (still) considered harmful. *IEEE CG&A*, 27(2):14–17, March 2007. doi: 10.1109/MCG.2007.323435

[16] M. Brehmer and T. Munzner. A multi-level typology of abstract visualization tasks. *IEEE TVCG*, 19(12):2376–2385, Dec 2013. doi: 10.1109/TVCG.2013.124

[17] K. Brett. Review of unclassed choropleth mapping. *Cartographic Perspectives*, 0(86), 2017. doi: 10.14714/CP86.1424

[18] S. Brooks and J. L. Whalley. A 2d/3d hybrid geographical information system. In *Proc. International Conference on Computer Graphics and Interactive Techniques in Australasia and South East Asia*, GRAPHITE '05, pp. 323–330. ACM, New York, 2005. doi: 10.1145/1101389.1101455

[19] H. Cleckner and T. Allen. Dasymetric mapping and spatial modeling of mosquito vector exposure, chesapeake, virginia, usa. *ISPRS international journal of geo-information*, 3(3):891–913, 2014. doi: 10.3390/ijgi3030891

[20] W. S. Cleveland and R. McGill. Graphical perception: Theory, experimentation, and application to the development of graphical methods. *Journal of the American Statistical Association*, 79(387):531–554, 1984. doi: 10.2307/2288400

[21] R. Coe. It's the effect size, stupid: What effect size is and why it is important. In *Annual Conf. British Educational Research Assoc.*, 2002.

[22] M. Correll, D. Moritz, and J. Heer. Value-suppressing uncertainty palettes. In *Proc. CHI*, pp. 642:1–642:11. ACM, New York, 2018. doi: 10.1145/3173574.3174216

[23] J. Crampton. GIS and geographic governance: reconstructing the choropleth map. *Cartographica*, 39(1):41–53, 2004. doi: 10.3138/H066-3346-R941-6382

[24] D. J. Cuff and K. R. Bieri. Ratios and absolute amounts conveyed by a stepped statistical surface. *The American Cartographer*, 6(2):157–168, 1979. doi: 10.1559/152304079784023087

[25] G. Cumming. The new statistics: Why and how. *Psychological Science*, 25(1):7–29, Jan. 2014. doi: 10.1177/0956797613504966

[26] J. Deber, R. Jota, C. Forlines, and D. Wigdor. How much faster is fast enough?: User perception of latency & latency improvements in direct and indirect touch. In *Proc. CHI*, pp. 1827–1836. ACM, New York, 2015. doi: 10.1145/2702123.2702300

[27] P. Deville, C. Linard, S. Martin, M. Gilbert, F. R. Stevens, A. E. Gaughan, V. D. Blondel, and A. J. Tatem. Dynamic population mapping using mobile phone data. *Proc. National Academy of Sciences*, 111(45):15888–15893, 2014.

[28] P. Dragicevic. Fair statistical communication in HCI. In J. Robertson and M. Kaptein, eds., *Modern Statistical Methods for HCI*, chap. 13, pp. 291–330. Springer International Publishing, Cham, Switzerland, 2016. doi: 10.1007/978-3-319-26633-6_13

[29] P. Dragicevic, F. Chevalier, and S. Huot. Running an HCI experiment in multiple parallel universes. In *CHI Extended Abstracts*, pp. 607–618. ACM, New York, 2014. doi: 10.1145/2559206.2578881

[30] R. Dunn. A dynamic approach to two-variable color mapping. *The American Statistician*, 43(4):245–252, 1989.

[31] C. L. Eicher and C. A. Brewer. Dasymetric mapping and areal interpolation: Implementation and evaluation. *Cartography and Geographic Information Science*, 28(2):125–138, 2001. doi: 10.1559/152304001782173727

[32] M. E. Elmer. Symbol considerations for bivariate thematic maps. In *Proc. International Cartographic Conference*, 2013.

[33] N. Elmqvist. *3D Occlusion Management and Causality Visualization*. Chalmers University of Technology, 2006.

[34] C. G. . T. Fernholz. Map: Donald Trump's "mean, mean, mean" health care bill is meanest to his most crucial voters.

[35] M. Friendly. Visions and re-visions of charles joseph minard. *Journal of Educational and Behavioral Statistics*, 27(1):31–51, 2002. doi: 10.3102/10769986027001031

[36] M. Friendly. The golden age of statistical graphics. *Statistical Science*, pp. 502–535, 2008. doi: 10.1214/08-STS26

[37] W. R. Garner. *The processing of information and structure.* Lawrence Erlbaum, Oxford, England, 1974.

[38] A. Gelman. No to inferential thresholds. Online. Visited 04/01/19, 2017.

[39] G. Gigerenzer. Mindless statistics. *The Journal of Socio-Economics*, 33(5):587–606, 2004. doi: 10.1016/j.socec.2004.09.033

[40] G. Gigerenzer. Statistical rituals: The replication delusion and how we got there. *Advances in Methods and Practices in Psychological Science*, p. 2515245918771329, 2018. doi: 10.1177%2F2515245918771329

[41] M. Gleicher, D. Albers, R. Walker, I. Jusufi, C. D. Hansen, and J. C. Roberts. Visual comparison for information visualization. *Information Visualization*, 10(4):289–309, 2011. doi: 10.1177/1473871611416549

[42] T. L. C. Griffin. Recognition of areal units on topological cartograms. *The American Cartographer*, 10(1):17–29, 1983. doi: 10.1559/152304083783948258

[43] R. Harris, M. Charlton, and C. Brunsdon. Mapping the changing residential geography of white british secondary school children in england using visually balanced cartograms and hexograms. *Journal of Maps*, 14(1):65–72, 01 2018. doi: 10.1080/17445647.2018.1478753

[44] R. Harris, M. Charlton, C. Brunsdon, and D. Manley. Balancing visibility and distortion: Remapping the results of the 2015 uk general election. *Environment and Planning A: Economy and Space*, 49(9):1945–1947, 2017. doi: 10.1177/0308518X17708439

[45] B. D. Hennig. Gridded cartograms as a method for visualising earthquake risk at the global scale. *Journal of Maps*, 10(2):186–194, 2014. doi: 10.1080/17445647.2013.806229

[46] D. H. House and C. J. Kocmoud. Continuous cartogram construction. In *Proc. VIS '98*, pp. 197–204, Oct 1998. doi: 10.1109/VISUAL.1998.745303

[47] Y. Jansen, P. Dragicevic, and J.-D. Fekete. Evaluating the Efficiency of Physical Visualizations. In *Proc. CHI*, pp. 2593–2602. ACM, ACM, Paris, France, Apr. 2013. doi: 10.1145/2470654.2481359

[48] L. Jégou. La troisième dimension en cartographie statistique, des cartes en prismes imprimées aux modèles 3d interactifs. *M@ppemonde*, 2007.

[49] G. F. Jenks. Generalization in statistical mapping. *Annals of the Association of American Geographers*, 53(1):15–26, 1963.

[50] G. F. Jenks and F. C. Caspall. Error on choroplethic maps: Definition, measurement, reduction. *Annals Association of American Geographers*, 61(2):217–244, 1971. doi: 10.1111/j.1467-8306.1971.tb00779.x

[51] B. Jürgen, M. Steiger, S. Mittelstädt, S. Thum, D. Keim, and J. Kohlhammer. A survey and task-based quality assessment of static 2d colormaps. In *Proc. SPIE*, 2015. doi: 10.1117/12.2079841

[52] S. Kaspar, S. I. Fabrikant, and P. Freckmann. Empirical study of cartograms. In *Proc. International Cartographic Conference*. International Cartographic Association, July 2011.

[53] D. Keating and L. Karklis. The increasingly diverse usa.

[54] M. Krzywinski and N. Altman. Points of significance: Error bars. *Nature Methods*, 10(10):921–922, Oct. 2013. doi: 10.1038/nmeth.2659

[55] P. Kubíček, M. Konečný, Z. Stachoň, J. Shen, L. Herman, T. Řezník, K. Staněk, R. Štampach, and Š. Leitgeb. Population distribution modelling at fine spatio-temporal scale based on mobile phone data. *International Journal of Digital Earth*, pp. 1–22, 2018.

[56] M. Le Goc, P. Dragicevic, S. Huron, J. Boy, and J.-D. Fekete. A Better Grasp on Pictures Under Glass: Comparing Touch and Tangible Object Manipulation using Physical Proxies. In *Proc. AVI*. Bari, Italy, June 2016. doi: 10.1145/2909132.2909247

[57] B. Lee, C. Plaisant, C. Sims, J.-D. Fekete, and N. Henry. Task taxonomy for graph visualization. In *Proc. BELIV '06*, pp. 1–5. ACM, ACM, Venezia, Italy, May 2006. doi: 10.1145/1168149.1168168

[58] E. Levasseur. La statistique graphique. *Journal of the Statistical Society of London*, pp. 218–250, 1885.

[59] C. T. Lloyd, H. Chamberlain, D. Kerr, G. Yetman, L. Pistolesi, F. R. Stevens, et al. Global spatio-temporally harmonised datasets for producing high-resolution gridded population distribution datasets. *Big Earth Data*, pp. 1–32, 2019. doi: 10.1080/20964471.2019.1625151

[60] L. R. Lucchesi and C. K. Wikle. Visualizing uncertainty in areal data with bivariate choropleth maps, map pixelation and glyph rotation. *Stat*, 6(1):292–302, 2017. doi: 10.1002/sta4.150

[61] A. Marcus, L. Feng, and J. I. Maletic. 3D representations for software vi-

sualization. In *Proc. ACM Symposium on Software Visualization*, SoftVis '03, pp. 27–ff. ACM, New York, 2003. doi: 10.1145/774833.774837

[62] A. M. Martinez-Graña, J. L. Goy, and C. Cimarra. 2D to 3D geologic mapping transformation using virtual globes and flight simulators and their applications in the analysis of geodiversity in natural areas. *Environ. Earth Sci.s*, 73(12):8023–8034, 2015. doi: 10.1007/s12665-014-3959-1

[63] A. Marín and M. Pelegrín. Towards unambiguous map labeling - integer programming approach and heuristic algorithm. *Expert Systems with Applications*, 98:221 – 241, 2018. doi: 10.1016/j.eswa.2017.11.014

[64] L. McNabb, R. S. Laramee, and R. Fry. Dynamic choropleth maps – using amalgamation to increase area perceivability. In *Proc. IV*, pp. 284–293, July 2018. doi: 10.1109/iV.2018.00056

[65] B. B. McShane and D. Gal. Statistical significance and the dichotomization of evidence. *Journal of the American Statistical Association*, 112(519):885–895, 2017. doi: 10.1080/01621459.2017.1289846

[66] B. Mericskay. La cartographie à l'heure du Géoweb : Retour sur les nouveaux modes de représentation spatiale des données numériques. *Cartes & géomatique*, 229-230:37–50, 2016.

[67] C. J. Minard. *Des tableaux graphiques et des cartes figuratives*. E. Thunot et Cie., 1861.

[68] C.-J. Minard. *La statistique*. Cusset et Company, 1869.

[69] M. Monmonier. How to lie with maps. *The American Statistician*, 51, 01 1996. doi: 10.2307/2685420

[70] G. D. Nelson and R. McKeon. Peaks of people: Using topographic prominence as a method for determining the ranked significance of population centers. *The Professional Geographer*, 71(2):342–354, 2019.

[71] R. K. Nelson, L. Winling, R. Marciano, N. Connolly, et al. Mapping inequality. *American panorama*, 2016.

[72] T. Niedomysl, E. Elldér, A. Larsson, M. Thelin, and B. Jansund. Learning benefits of using 2d versus 3d maps: Evidence from a randomized controlled experiment. *Journal of Geography*, 112(3):87–96, 2013. doi: 10.1080/00221341.2012.709876

[73] S. Nusrat, M. J. Alam, and S. Kobourov. Evaluating cartogram effectiveness. *IEEE TVCG*, 24(2):1077–1090, Feb 2018. doi: 10.1109/TVCG.2016.2642109

[74] S. Nusrat and S. Kobourov. Visualizing cartograms: Goals and task taxonomy. *arXiv preprint arXiv:1502.07792*, 2015.

[75] L. Padilla, P. S. Quinan, M. Meyer, and S. H. Creem-Regehr. Evaluating the impact of binning 2d scalar fields. *IEEE TVCG*, 23(1):431–440, Jan 2017. doi: 10.1109/TVCG.2016.2599106

[76] G. Palsky. *Des chiffres et des cartes: naissance et développement de la cartographie quantitative française au XIXe siecle*, vol. 19. Comité des travaux historiques et scientifiques-CTHS, 1996.

[77] G. Palsky. Paris en chiffres les premiers atlas statistiques de paris. *Le Monde des cartes*, (171):52–58, 2002.

[78] F. Pappenberger, H. L. Cloke, and C. A. Baugh. Cartograms for use in forecasting weather-driven natural hazards. *The Cartographic Journal*, pp. 1–12, 02 2019. doi: 10.1080/00087041.2018.1534358

[79] R. Parrott, S. Hopfer, C. Ghetian, and E. Lengerich. Mapping as a visual health communication tool: Promises and dilemmas. *Health Communication*, 22(1):13–24, 2007. doi: 10.1080/10410230701310265

[80] M. P. Peterson. An evaluation of unclassed crossed-line choropleth mapping. *The American Cartographer*, 6(1):21–37, 1979. doi: 10.1559/152304079784022736

[81] W. J. Requia, P. Koutrakis, and A. Arain. Modeling spatial distribution of population for environmental epidemiological studies: Comparing the exposure estimates using choropleth versus dasymetric mapping. *Environment International*, 119:152 – 164, 2018. doi: 10.1016/j.envint.2018.06.021

[82] P. Rheingans. Dynamic color mapping of bivariate qualitative data. In *Proc. VIS '97*, pp. 159–166, Oct 1997. doi: 10.1109/VISUAL.1997.663874

[83] P. K. Robertson and J. F. O'Callaghan. The generation of color sequences for univariate and bivariate mapping. *IEEE CG&A*, 6(2):24–32, Feb 1986. doi: 10.1109/MCG.1986.276688

[84] B. E. Rogowitz and A. D. Kalvin. The "which blair project": a quick visual method for evaluating perceptual color maps. In *Proc. VIS '01.*, pp. 183–556, Oct 2001. doi: 10.1109/VISUAL.2001.964510

[85] B. E. Rogowitz, L. A. Treinish, and S. Bryson. How not to lie with visualization. *Computers in Physics*, 10(3):268–273, 1996. doi: 10.1063/1.4822401

[86] R. E. Roth. An empirically-derived taxonomy of interaction primitives for interactive cartography and geovisualization. *IEEE TVCG*, 19(12):2356–2365, 2013. doi: 10.1109/TVCG.2013.130

[87] R. E. Roth. *Visual Variables*, pp. 1–11. American Cancer Society, 2017. doi: 10.1002/9781118786352.wbieg0761

[88] R. E. Roth, A. W. Woodruff, and Z. F. Johnson. Value-by-alpha maps: An alternative technique to the cartogram. *The Cartographic Journal*, 47(2):130–140, 2010. doi: 10.1179/000870409X12488753453372

[89] M. A. Rylov and A. W. Reimer. A comprehensive multi-criteria model for high cartographic quality point-feature label placement. *Cartographica*, 49(1):52–68, 2014. doi: 10.3138/carto.49.1.2137

[90] J. Sauro and J. R. Lewis. Average task times in usability tests: What to report? In *Proc. CHI*, pp. 2347–2350. ACM, New York, 2010. doi: 10.1145/1753326.1753679

[91] V. Shandas, J. Voelkel, M. Rao, and L. George. Integrating high-resolution datasets to target mitigation efforts for improving air quality and public health in urban neighborhoods. *Int. J. Environ. Res. Public Health*, 13(8), 2016. doi: 10.3390/ijerph13080790

[92] B. Shneiderman. The eyes have it: a task by data type taxonomy for information visualizations. In *Proc. IEEE Symposium on Visual Languages*, pp. 336–343, Sep. 1996. doi: 10.1109/VL.1996.545307

[93] Y. B. Shrinivasan and J. J. van Wijk. Supporting the analytical reasoning process in information visualization. In *Proc. CHI*, pp. 1237–1246. ACM, New York, 2008. doi: 10.1145/1357054.1357247

[94] B. Speckmann and K. Verbeek. Necklace maps. *IEEE TVCG*, (6):881–889, 2010. doi: 10.1109/TVCG.2010.180

[95] J. Stewart and P. J. Kennelly. Illuminated choropleth maps. *Annals of the Association of American Geographers*, 100(3):513–534, 2010. doi: 10.1080/00045608.2010.485449

[96] H. Sun and Z. Li. Effectiveness of cartogram for the representation of spatial data. *The Cartographic Journal*, 47(1):12–21, 2010. doi: 10.1179/000870409X12525737905169

[97] M. Tao. *Using cartograms in disease mapping*. PhD thesis, University of Sheffield, Department of Geography, 2010.

[98] W. Tobler. Thirty five years of computer cartograms. *Association of American Geographers*, 94(1):58–73, 2004. doi: 10.1111/j.1467-8306.2004.09401004.x

[99] W. R. Tobler. Choropleth maps without class intervals? *Geographical Analysis*, 5(3):262–265, 1973. doi: 10.1111/j.1538-4632.1973.tb01012.x

[100] B. E. Trumbo. A theory for coloring bivariate statistical maps. *The American Statistician*, 35(4):220–226, 1981.

[101] G. R. VandenBos, ed. *Publication Manual of the American Psychological Association*. American Psychological Association, 6th ed., 2009.

[102] H. Wainer and C. M. Francolini. An empirical inquiry concerning human understanding of two-variable color maps. *The American Statistician*, 34(2):81–93, 1980.

[103] X. Wang, L. Besançon, M. Ammi, and T. Isenberg. Augmenting tactile 3D data navigation with pressure sensing. *Computer Graphics Forum*, 38(3), June 2019. doi: 10.1111/cgf.13716

[104] C. Ware. Color sequences for univariate maps: Theory, experiments and principles. *IEEE CG&A*, 8(5):41–49, Sep. 1988. doi: 10.1109/38.7760

[105] C. Ware. Quantitative texton sequences for legible bivariate maps. *IEEE TVCG*, 15(6):1523–1530, Nov 2009. doi: 10.1109/TVCG.2009.175

[106] J. A. Ware. Using animation to improve the communicative aspect of cartograms. Master's thesis, 1999.

[107] A. Williams and A. Emamdjomeh. America is more diverse than ever — but still segregated.

[108] Y. Yao, X. Liu, X. Li, J. Zhang, Z. Liang, K. Mai, and Y. Zhang. Mapping fine-scale population distributions at the building level by integrating multisource geospatial big data. *INT J GEOGR INF SCI*, 31(6):1220–1244, 2017.

[109] J. S. Yi, Y. a. Kang, J. Stasko, and J. Jacko. Toward a deeper understanding of the role of interaction in information visualization. *IEEE TVCG*, 13(6):1224–1231, Nov. 2007. doi: 10.1109/TVCG.2007.70515

[110] L. Yu, K. Efstathiou, P. Isenberg, and T. Isenberg. CAST: Effective and efficient user interaction for context-aware selection in 3D particle clouds. *IEEE TVCG*, 22(1):886–895, Jan. 2016. doi: 10.1109/TVCG.2015.2467202

[111] S. Zhai, W. Buxton, and P. Milgram. The partial-occlusion effect: Utilizing semitransparency in 3d human-computer interaction. *ACM ToCHI*, 3(3):254–284, Sept. 1996. doi: 10.1145/234526.234532

[112] M. X. Zhou and S. K. Feiner. Visual task characterization for automated visual discourse synthesis. In *Proc. CHI*, pp. 392–399. ACM, New York, 1998. doi: 10.1145/274644.274698

[113] H. Zoraghein and S. Leyk. Data-enriched interpolation for temporally consistent population compositions. *GIScience & Remote Sensing*, 56(3):430–461, 2019. doi: 10.1080/15481603.2018.1509463